\documentclass[aps,prc,twocolumn,10pt, superscriptaddress, showpacs, floatfix]{revtex4-1}

\usepackage{amssymb,epsfig}
\usepackage{bbold}

\newcommand{\be}{\begin{equation}}
\newcommand{\ee}{\end{equation}}
\newcommand{\bea}{\begin{eqnarray}}
\newcommand{\eea}{\end{eqnarray}}

\begin{document}

\title{Fragmentation patterns of nuclear response: low-spin giant resonances and soft modes}
\author{Elena Litvinova}
\affiliation{Department of Physics, Western Michigan University, Kalamazoo, MI 49008, USA}
\affiliation{Facility for Rare Isotope Beams, Michigan State University, East Lansing, MI 48824, USA}
\affiliation{GANIL, CEA/DRF-CNRS/IN2P3, F-14076 Caen, France}

\date{\today}

\begin{abstract}
Nuclear resonances provide a rich and versatile testbed for exploring fundamental aspects of physics, particularly within the domain of strongly correlated many-body systems. The overarching goal of the theory is to develop a consistent and predictive framework that is (i) capable of a spectroscopically accurate description and (ii) sufficiently general to be applied across different energy scales and transferable to a wide range of complex systems. Thoroughly capturing emergent collective phenomena that arise in nuclear media is the central challenge for the theory, which is discussed in this contribution.
It concentrates on the themes inspired and influenced by Angela Bracco's research, in particular, on the fragmentation patterns of the monopole and dipole responses of medium-heavy nuclei and associated open problems. 
\end{abstract}

\maketitle


\section{Introduction} 

The theoretical description of nuclear giant resonances and soft modes is one of the most complex areas of nuclear structure physics. 
These excitations probe various facets of nuclear forces and strong many-body correlations, and despite decades-long ongoing effort, there is no satisfactory theoretical description of experimental data across the nuclear landscape. 

The intricate nature of the nucleon-nucleon (NN) interaction introduces limitations on its precise knowledge. The NN interaction is definition-sensitive regardless of whether it is extracted from data on NN scattering or computed from quantum chromodynamics. Various approximations compatible with the symmetry constraints can be applied to its analytical form, but there is a general consensus that the NN interaction is used as an input for the nuclear many-body theory in the form of a static potential, although its validity remains an open question. Furthermore, accurate solutions to the strongly correlated many-body problem are hardly tractable computationally, independently of the interaction used.
Addressing these challenges in applications to medium-mass and heavy nuclei requires advanced, non-perturbative techniques based on quantum field theory, such as the Green function method, equations of motion, and cluster decomposition strategies, tailored to specific nuclear structure problems.


Quantifying nuclear structure with Green functions (GFs), corresponding to nucleonic propagators, offers an efficient and transferable approach as the GFs or, more generally, correlation functions (CFs), connect interacting quantum many-body systems across energy scales. The GF's direct relation to observables facilitates identifying emergent phenomena and universal spectral patterns \cite{Matsubara1955,Watson1956,Brueckner1955,Martin1959,Ethofer1969,SchuckEthofer1973}. For instance, in atomic nuclei, the single-nucleon propagators contain odd-particle energies and occupancies, benchmarked by transfer or knock-out reactions, while the two-time particle-hole propagators describe the nuclear response to external probes. Superfluid pairing correlations manifest through pair propagators, whose residues are related to pairing gaps \cite{Gorkov1958,Kadanoff1961}.
More complex CFs, such as three-fermion and four-fermion ones, enter the theory via the interaction kernels of equations of motion (EOMs) for the lower-rank GFs \cite{Martin1959,Ethofer1969,SchuckEthofer1973}. These higher-rank structures, encoding multi-fermion correlations, are crucial for capturing long-range dynamics and feedback effects on short-range physics. In medium and strongly coupled regimes, like in atomic nuclei, they underlie emergent collective modes, enabling the identification of order parameters and the mapping of elementary to emergent degrees of freedom.

Traditional GF formulations use non-symmetric dynamical kernels \cite{Migdal1967}, often treated via factorizations into one-fermion CFs. Peter Schuck and collaborators emphasized the advantage of symmetric forms of these kernels for tractable approximations via cluster decompositions \cite{AdachiSchuck1989,Danielewicz1994,DukelskyRoepkeSchuck1998}. This method, capable of retaining exact two-fermion CFs, bridges nuclear physics, condensed matter, and quantum chemistry \cite{Tiago2008,Sangalli2011,SchuckTohyama2016,Popovici2011,LitvinovaSchuck2019,Litvinova2023a}. Such approximations derived from bare Hamiltonians can be connected to beyond-mean-field (BMF) approaches, historically using effective interactions  \cite{Schuck2019,LitvinovaSchuck2020}. In the approaches to the nuclear response going beyond the basic random phase approximation (RPA),
the emergent collectivity, crucial for the description of medium-heavy nuclei, can be modeled by retaining correlated two-quasiparticle (phonon) pairs in the EOM kernel. This idea, central to the nuclear field theory (NFT)\cite{BohrMottelson1969,BohrMottelson1975,Broglia1976,BortignonBrogliaBesEtAl1977,BertschBortignonBroglia1983,Barranco2017}, underlies (quasi)particle-vibration coupling (qPVC) schemes, also featured in the extensions of Migdal's theory \cite{KamerdzhievSpethTertychny2004,Tselyaev1989,LitvinovaTselyaev2007} and the quasiparticle-phonon model (QPM) \cite{Soloviev1992,Malov1976,Ponomarev1999,Lenske:2019ubp,LoIudice2012,Andreozzi2008,Knapp:2014xja,Knapp:2015wpt,DeGregorio2016,DeGregorio2016a}. Here, the dynamical kernels account for long-range effects through fully correlated four-fermion CFs. While exact solutions are intractable, factorized dynamical kernels and tailored approximations to their static counterparts (e.g., via interactions fitted within the density functional theory (DFT) on the mean-field level or unitary-transformed bare interactions) enable self-consistent implementations, see, for instance, Refs. \cite{Drozdz:1990zz,PapakonstantinouRoth2009,GambacurtaGrassoCatara2010,Gambacurta2015} employing non-superfluid second RPA  with pure two-particle-two-hole ($2p2h$) configurations and Refs. \cite{LitvinovaRingTselyaev2008,LitvinovaRingTselyaev2010,Tselyaev2013,RobinLitvinova2016,Tselyaev2018,Robin2019,Li2022,Litvinova2023} with qPVC and superfluidity included. 
 The latter approaches highlight the importance of correlated $2p2h$ configurations beyond RPA, which are taken into account via qPVC. The $2p2h$ configurations induce fragmentation of the giant resonances and soft modes with considerable improvements in their description.
 Yet, accurate spectroscopy demands inclusion of higher complexity. QPM introduced multiphonon wavefunctions (up to $3p3h$) \cite{Ponomarev1999,Savran2011,Lenske:2019ubp}, though implementations are limited by non-self-consistency and parameter tuning. Recent advances enable handling $3p3h$ configurations over broad energy ranges \cite{LitvinovaSchuck2019,Novak2024,Muescher2024} in a relativistic framework. 
Possessing a systematic character and convergence, the approach of progressive inclusion of complex configurations $npnh$ with growing degree of complexity $n$ on one hand, enables uncertainty quantification. On the other hand, it introduces new uncertainties stemming from the choice of a calculation scheme, classifying nucleonic correlations, and adjusting the parameters of effective interaction in both $ph$ and $pp$ channels. The importance of pairing gap parametrization was addressed, e.g., in Ref. \cite{Markova2025}, quantifying the sensitivity of the pygmy dipole resonance to the pairing gap value, which occurs already in the leading-order qPVC approach. This and other facets of complex configurations, appearing in the practical implementations, call for further clarification and establishing their links to the model uncertainties.       
 
In this context, I focus here on the leading qPVC mechanism of fragmentation of nuclear resonances and discuss recent progress in the description of the monopole and dipole modes, particularly the aspects of qPVC that make it difficult to achieve spectroscopic accuracy. 

%



\section{Nuclear response theory in a nutshell}
\label{response}

The nuclear response theory is the most convenient tool to quantify the nuclear spectra, providing both the excitation energies and transition probabilities across the energy regimes in terms of the strength functions (SFs). The collection of SFs with various angular momentum, spin, and isospin transfer contains all the information about nuclear structure, which can be accessed via the probes associated with one-body operators.  

For sufficiently weak external fields, the SF is defined by Fermi's golden rule:
\bea
S_F(\omega) = \sum\limits_{\nu>0} \Bigl[ |\langle \nu|F^{\dagger}|0\rangle |^2\delta(\omega-\omega_{\nu}) - |\langle \nu|F|0\rangle |^2\delta(\omega+\omega_{\nu})
\Bigr], \nonumber\\
\label{SF}
\eea
where the index $\nu$ in the summation runs over the excited states $|\nu\rangle$ accessible by the given probe described by the quantum mechanical operator $F^{\dagger}$ acting on the ground state $|0\rangle$. The matrix element $\langle \nu|F^{\dagger}|0\rangle$ is the transition amplitude, and for the typical one-body operator reads,
\be
\langle \nu|F^{\dagger}|0\rangle = \sum\limits_{12}\langle \nu|F_{12}^{\ast}\psi^{\dagger}_2\psi_1|0\rangle = \sum\limits_{12}F_{12}^{\ast}\rho_{21}^{\nu\ast},
\label{Frho}
\ee
where the transition densities $\rho^{\nu}$ are introduced as
\be
\rho^{\nu}_{12} = \langle 0|\psi^{\dagger}_2\psi_1|\nu \rangle. 
\label{trden}
\ee
The physical meaning of the transition density (\ref{trden}) is the overlap between the pure particle-hole configuration $\psi^{\dagger}_2\psi_1$, built on top of the ground state $|0\rangle$, and the formally exact, or correlated, excited state $|\nu\rangle$. Here, $\psi_1$ and $\psi_1^{\dagger}$ are the nucleonic field operators in the basis $\{1\}$, where the number index comprises all the quantum numbers completely quantifying a single-particle state. This can be the set of coordinates or momentum, spin, and isospin, or a basis orbital in a confining potential or a self-consistent mean field. 
Conventionally, the delta functions in Eq. (\ref{SF}) can be represented by the Lorentz distribution with an infinitesimal width $\Delta$,
so that
\be
S_F(\omega) 
= -\frac{1}{\pi}\lim\limits_{\Delta \to 0} {\Im} \Pi(\omega+ i\Delta),
\label{SFDelta} 
\ee
relating the nuclear polarizability $\Pi(\omega)$
\be
\Pi(\omega) 
=  \sum\limits_{\nu} \Bigl[ \frac{B_{\nu}}{\omega - \omega_{\nu}}
- \frac{{\bar B}_{\nu}}{\omega + \omega_{\nu} }
\Bigr]
\label{Polar}
\ee
to the transition probabilities $B_{\nu}$ and ${\bar B}_{\nu}$
\be
B_{\nu} = |\langle \nu|F^{\dagger}|0\rangle |^2\ \ \ \ \ \ \ \ 
{\bar B}_{\nu} = |\langle \nu|F|0\rangle |^2 ,
\label{Prob}
\ee
 corresponding to absorption and emission, respectively.
The strength, or spectral, function for the nuclear response to the probe $F$ reads:
\be
S_F(\omega) = -\frac{1}{\pi}\lim_{\Delta\to 0}\Im\sum\limits_{121'2'}F_{12}R_{12,1'2'}(\omega+i\Delta)F^{\ast}_{1'2'}.
\label{SFF}
\ee
In practical implementations, $\Delta$ becomes a finite parameter, which serves for (i) representation of the delta-function by a finite-width distribution, (ii) imitation of the non-zero dynamical self-energy or part of it not taken into account explicitly, (iii) imitation of the continuum width, known from direct continuum calculations \cite{Kamerdzhiev1998,DaoutidisRing2011,LitvinovaBelov2013}, in approaches with "discretized" continuum, and (iv) adjusting $S_F(\omega)$ to variability of experimental resolution or binning. The in-medium physics of nuclear structure is, thereby, delegated to the response function $R_{12,1'2'}(\omega)$:
\bea
R_{12,1'2'}(\omega) = \sum\limits_{\nu>0}\Bigl[ \frac{\rho^{\nu}_{21}\rho^{\nu\ast}_{2'1'}}{\omega - \omega_{\nu} + i\delta} -  \frac{\rho^{\nu\ast}_{12}\rho^{\nu}_{1'2'}}{\omega + \omega_{\nu} - i\delta}\Bigr], \nonumber\\
\delta \to +0 .
\label{respspec}
\eea
It is clear from Eqs. (\ref{SFF}) and (\ref{respspec}) that the poles of $R_{12,1'2'}(\omega)$ at the energies $\omega_{\nu} = E_{\nu} - E_0$ of the excited states with respect to the ground state energy become the locations of the peaks in the strength function, while the residues of $R_{12,1'2'}(\omega)$ are connected to the peak heights as
\be
B_{\nu} = \pi \Delta S_F(\omega_{\nu}).
\ee
Eq. (\ref{respspec}) is linked to the field theoretical formalism, namely, it is the Fourier image of the particle-hole propagator on the background of the interacting ground state:
\be
R_{12,1'2'}(t-t') =  -i\langle T\psi^{\dagger}(1)\psi(2)\psi^{\dagger}(2')\psi(1')\rangle,
\label{phresp}
\ee
where $\langle...\rangle$ is a shorthand notation for the ground state expectation value and $\psi(1), \psi^{\dagger}(1)$ are the Heisenberg fermionic field operators. 
Strictly speaking, the complete Fourier transform contains the $\nu = 0$, i.e., the ground state term, which is typically excluded from the response function \cite{Dickhoff2005}. 

\begin{figure*}
\begin{center}
\includegraphics[width=0.9\textwidth]{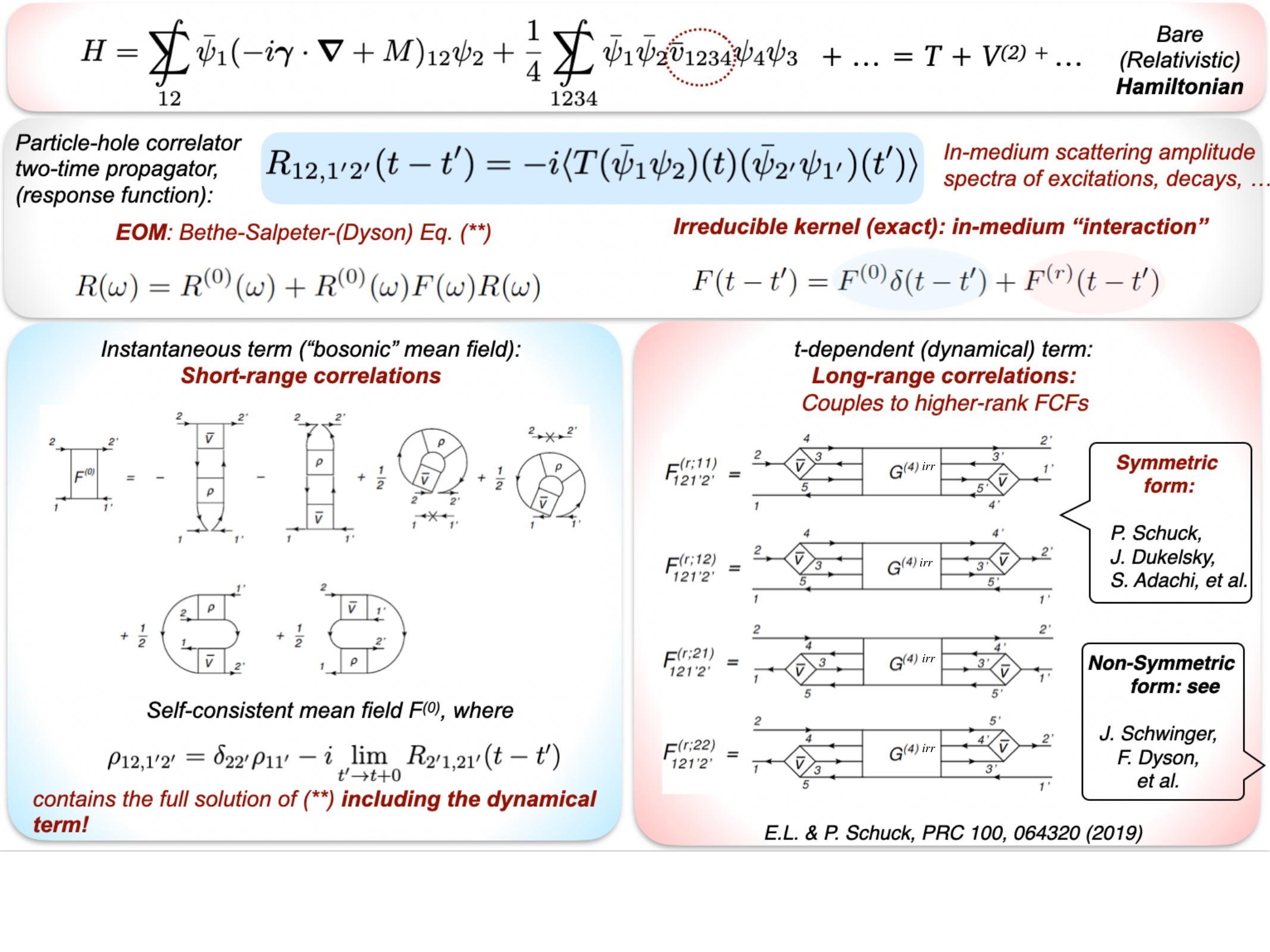}
\end{center}
\vspace{-1cm}
\caption{Schematic structure of the fermionic response theory. The rectangular blocks marked with $\rho$ stand for exact two-fermion densities, and those with $G^{(4)}$ denote correlated four-fermion, or $2p2h$, GF.}
\label{Response}%
\end{figure*}

The resulting spectrum is a sum of, in general, overlapping peaks, whose number is equal to the number of terms in Eq. (\ref{respspec}).  This number, as well as the peak locations and heights, depend on the correlation content of the theory, which defines the interaction kernel of the EOM for $R_{12,1'2'}$.  This EOM is essentially the Bethe-Salpeter equation, which acquires the Dyson form because of the definition (\ref{phresp}) as a manifestly two-time correlation function. That is, we operate the Bethe-Salpeter-Dyson equation (BSDE), which, in the symbolic form in the energy domain, reads:
\be
R(\omega) = R^0(\omega) + R^0(\omega)\Bigl( F^0 + F^r(\omega)\Bigr)R(\omega).
\label{BSDE}
\ee
Here, $R^0(\omega)$ is the non-interacting particle-hole propagator and the interaction kernel $F(\omega) = F^0 + F^r(\omega)$ absorbs all the in-medium physical processes. Note that at this point the theory is exact, and the decomposition of $F(\omega)$ into the static $F^0$ and dynamical  $F^r(\omega)$ terms is of a fundamental character, reflecting the non-time-dependent nature of the nucleon-nucleon (NN) interaction used from the starting Hamiltonian
\be
H =  \sum_{12}h_{12}\psi^{\dag}_1\psi_2 + \frac{1}{4}\sum\limits_{1234}{\bar v}_{1234}{\psi^{\dagger}}_1{\psi^{\dagger}}_2\psi_4\psi_3.
\label{Hamiltonian1}
\ee
It is completely determined by its one-body kinetic $h_{12}$ and two-body NN interaction ${\bar v}_{1234}$ antisymmetrized matrix elements. At this point, we note that higher-body forces are omitted in the Hamiltonian until their necessity is better justified. Thereby, in principle, the only input of the theory is the nucleonic masses, intrinsic spins, and their pairwise interaction in the vacuum $v$. The diagrammatic structure of the general exact interaction kernels is sketched in Fig. \ref{Response}.

\begin{figure*}
\begin{center}
\includegraphics[width=0.9\textwidth]{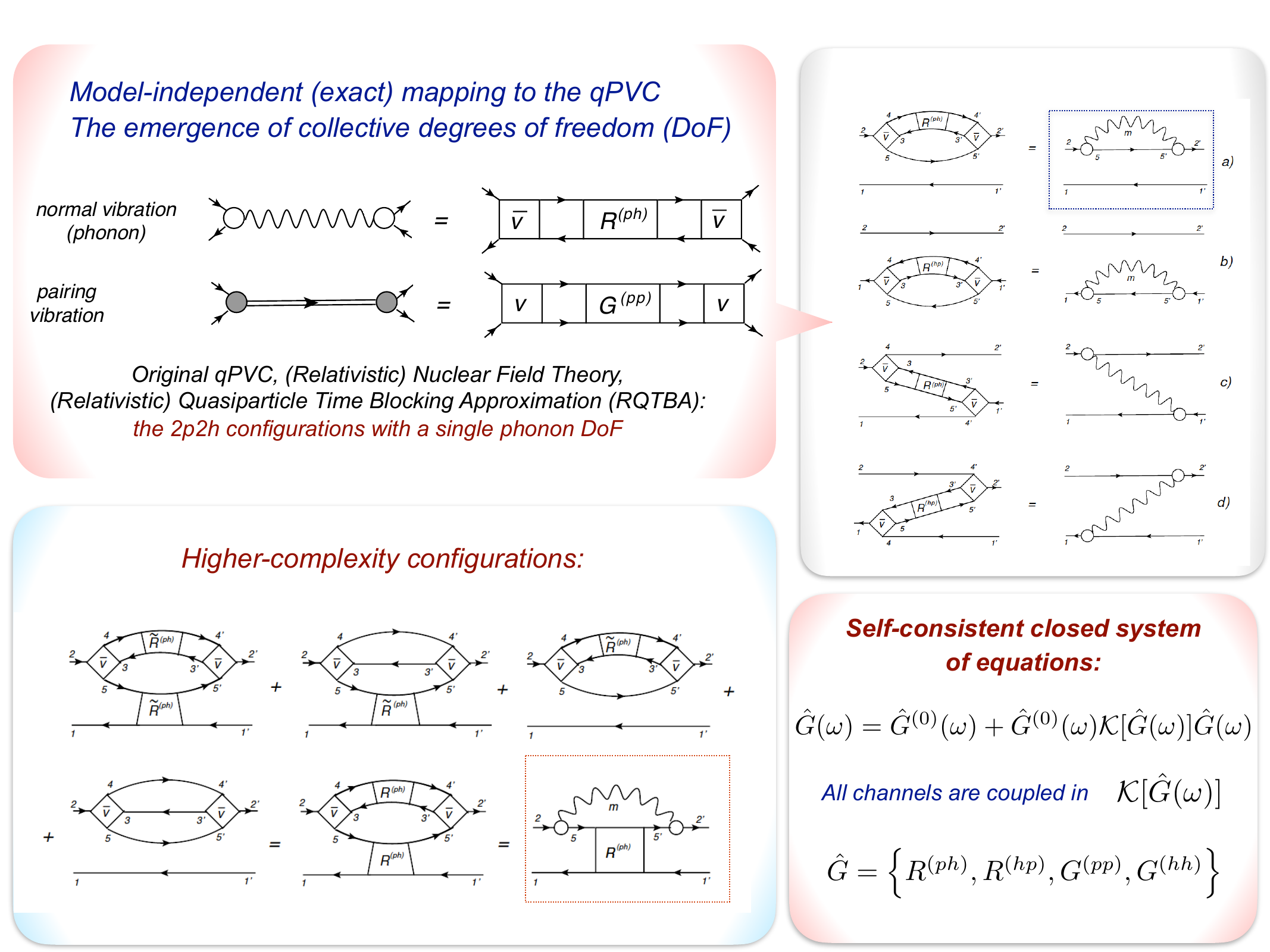}
\end{center}
\caption{Schematic structure of the fermionic response theory, truncated on the two-body level via the cluster decomposition and mapped to qPVC. $\tilde R$ stands for the correlated part of the response function.}
\label{Mapping}%
\end{figure*}

Thereby, the exact response theory is well defined; however, in practice, neither $F^0$ nor $F^r(\omega)$ is calculable. $F^0$ consists of ${\bar v}_{1234}$ matrix elements contracted with the two-fermion densities, related to the BSDE solutions, and $F^r$ contains a higher-rank, namely $2p2h$, exact correlation function $G^{(4)}$, which requires a solution of its own EOM, coupled to even higher-rank correlation functions. Various approximations may be derived via a cluster decomposition of $G^{(4)}$:
\bea
G^{(4q)irr} \sim G^{(q)}G^{(q)}G^{(q)}G^{(q)} + G^{(q)}G^{(q)}G^{(2q)} \nonumber \\
+ G^{(2q)}G^{(2q)} + G^{(q)}G^{(3q)} + \sigma^{(4q)},
\label{CD2}
\eea
where the index '$q$' (quasiparticle) covers both particles '$p$' and holes '$h$', and the symbol $G^{(nq)}$ stands for any type of correlation functions with $n$ quasiparticle operators. Furthermore, Bogoliubov's transformation enables the formulation of the response theory for superfluid systems in the complete analogy to the non-superfluid one, with the replacement of particles and holes by quasiparticles  \cite{Litvinova2022}.
The qPVC approaches are associated with retaining the terms with at least one $G^{(2q)}$ in Eq. (\ref{CD2}), which, by contraction with the pairs of interaction matrix elements, can be exactly mapped to the qPVC amplitudes as schematically shown in Fig. \ref{Mapping}. 

%


On the most basic level, the response theory is the random phase approximation (RPA) or its superfluid extension, quasiparticle RPA (QRPA), typically associated with the linear response \cite{RingSchuck1980}. In the diagrammatic language, (Q)RPA is given by a one-loop diagram of the two-fermion propagator in a correlated medium in the limit of vanishing $2p2h$ and more complex correlations in the interaction kernel of the BSDE, i.e., simply setting $F^{r}(\omega) = 0$ in Eq. (\ref{BSDE}).  In Rowe's EOM, (Q)RPA is generated by the one-particle-one-hole (two-quasiparticle) $1p1h$ (2q) excitation operator \cite{Rowe1968}, creating excited states by acting on a Hartree-Fock (Hartree-Fock-Bogoliubov) ground state, also processed through an EOM.

In practical implementations for atomic nuclei, (Q)RPA is known to satisfactorily describe the general properties of giant resonances (GRs), such as their centroids and sum rules. The widths of GRs and gross structures are poorly captured by (Q)RPA in both spherical and deformed nuclei, although a bit better in the latter case.
Soft modes below the giant resonances are more sensitive to the dynamical correlations, which are missing in (Q)RPA, so that at low energies this approach is unable to provide the richness of the observed fine spectral details. More accurate solutions involve higher complexity ($npnh$) correlations in both excited and ground states of the nucleus. Approximations beyond (Q)RPA follow from the BSDE, retaining the dynamical kernel. The leading approximation to it is the minimal coupling scheme, which includes $2q\otimes phonon$ configurations, dubbed as the leading-order qPVC. The phonons emerge from the cluster decomposition of the exact dynamical kernel as correlated $2q$ pairs, as illustrated in Fig. \ref{Mapping}, quantified by the qPVC vertices and vibrational frequencies. The qPVC vertices are then interpreted as the new order parameters that can be used for constructing systematic expansions and iterative non-perturbative solutions. Essentially, in these variants of the cluster decomposition, the response theory acquires a closed form, as outlined in the lower right panel of Fig. \ref{Mapping}.

The static kernel of the response EOM is also problematic for practical implementations, and so far, there have been no realizations of $F^0$ with the bare interaction and satisfactorily calculated two-body densities. While this remains a task for future efforts, approximating $F^0$ by effective interactions adjusted in the DFT framework provides an optimal strategy. With such interactions, reasonable phonon characteristics for quantifying the dynamical kernel can be obtained already within (Q)RPA. Then, $F^r(\omega)$ approximated by qPVC with varying degrees of complexity can be corrected by a subtraction, restoring the self-consistency of the 
ab-initio framework \cite{Tselyaev2013}. The main purpose of the subtraction is to eliminate double-counting of qPVC, which arises when an effective interaction adjusted at the mean-field level is used. In this case, the entire interaction kernel shown in Fig. \ref{Response} is approximated by a single static term as $F^0 + F^r(\omega) \approx {\tilde F}^0$, i.e., all the loop contributions are absorbed into the tree-level term(s) with refitted parameters. When the dynamical term is added back in a beyond-mean-field approach, a double counting occurs, which is then corrected by subtracting the contribution $F^r(\omega = 0)$. Besides taking care of the unwanted double-counting, in practice, this subtraction also ensures a better convergence of the qPVC loop contributions detailed in Fig. \ref{Mapping} and preserves the placement of the spurious states. More details can be found in the original work \cite{Tselyaev2013}. Note that by "ab initio" above and below, I mean that the only input to the theory is interaction between two particles in the vacuum, as it is commonly adopted in theoretical physics. 

The first self-consistent microscopic approach with qPVC in terms of $2q\otimes phonon$ configurations (the NFT dynamical kernel given in the upper right panel of Fig. \ref{Mapping}, generalized to the superfluid phase) was realised in Ref. \cite{LitvinovaRingTselyaev2008}.  This implementation, applied to the dipole response of medium-heavy nuclei, was a step toward a universal theory of nuclear structure. The response EOM then keeps the algebraic structure of the ab-initio BSDE and is rooted in particle physics by employing the effective meson-exchange interaction \cite{Lalazissis1997,NL3star}. This early approach was based on the phenomenological assumption about the leading role of $2q\otimes phonon$ configurations and derived with the time blocking technique, following the non-relativistic non-superfluid approach of Ref. \cite{Tselyaev1989}, therefore, it was identified as relativistic time blocking approximation (RQTBA). In Refs. \cite{LitvinovaSchuck2019,Litvinova2022}, the more general response theory was obtained via ab-initio EOMs, where both the phenomenological qPVC and time blocking were ruled out as unnecessary ingredients.  
The relativistic EOM with $2q\otimes phonon$ (REOM$^2$) configurations obtained with the relativistic QRPA (RQRPA) phonons was shown to be equivalent to RQTBA in the resonant approximation, neglecting qPVC-associated ground state correlations. The major breakthrough was overcoming the limitations of the phenomenological approach and extending the dynamical kernel to configurations of arbitrary complexity. 
An example of such an extension was presented as REOM$^3$ accommodating $2q\otimes 2phonon$ configurations in Refs. \cite{LitvinovaSchuck2019,Litvinova2023a,Novak2024} by reprocessing the response function in the dynamical kernel as outlined in the lower panels of Fig. \ref{Mapping}.

\section{Nuclear response in the relativistic NFT (RNFT) framework}
\label{calculations}

\subsection{Isoscalar monopole response}

The theoretical description of the nuclear isoscalar monopole response deals with the operator \cite{GR2001,Garg2018}:
\be
F_{00} = \sum\limits_{i = 1}^Ar_i^2,
\ee 
where $A$ is the nuclear mass number. This type of response is associated predominantly with compression (breathing) modes, probed experimentally by the inelastic scattering of alpha particles or deuterons to small forward angles. The strength distribution in medium-heavy nuclei exhibits a pronounced peak at 15-20 MeV, named the isoscalar giant monopole resonance (ISGMR). It is related to small density fluctuations around the saturation point of the nuclear matter and allows for extracting
the nuclear (in)compressibility coefficient $K_{\infty}$. A direct link established between the centroid of the ISGMR and $K_{\infty}$, 
makes the ISGMR studies the main source of fundamental information about the nuclear equation of state.  

\begin{figure}
\begin{center}
\vspace{-0.5cm}
\includegraphics[width=0.55\textwidth]{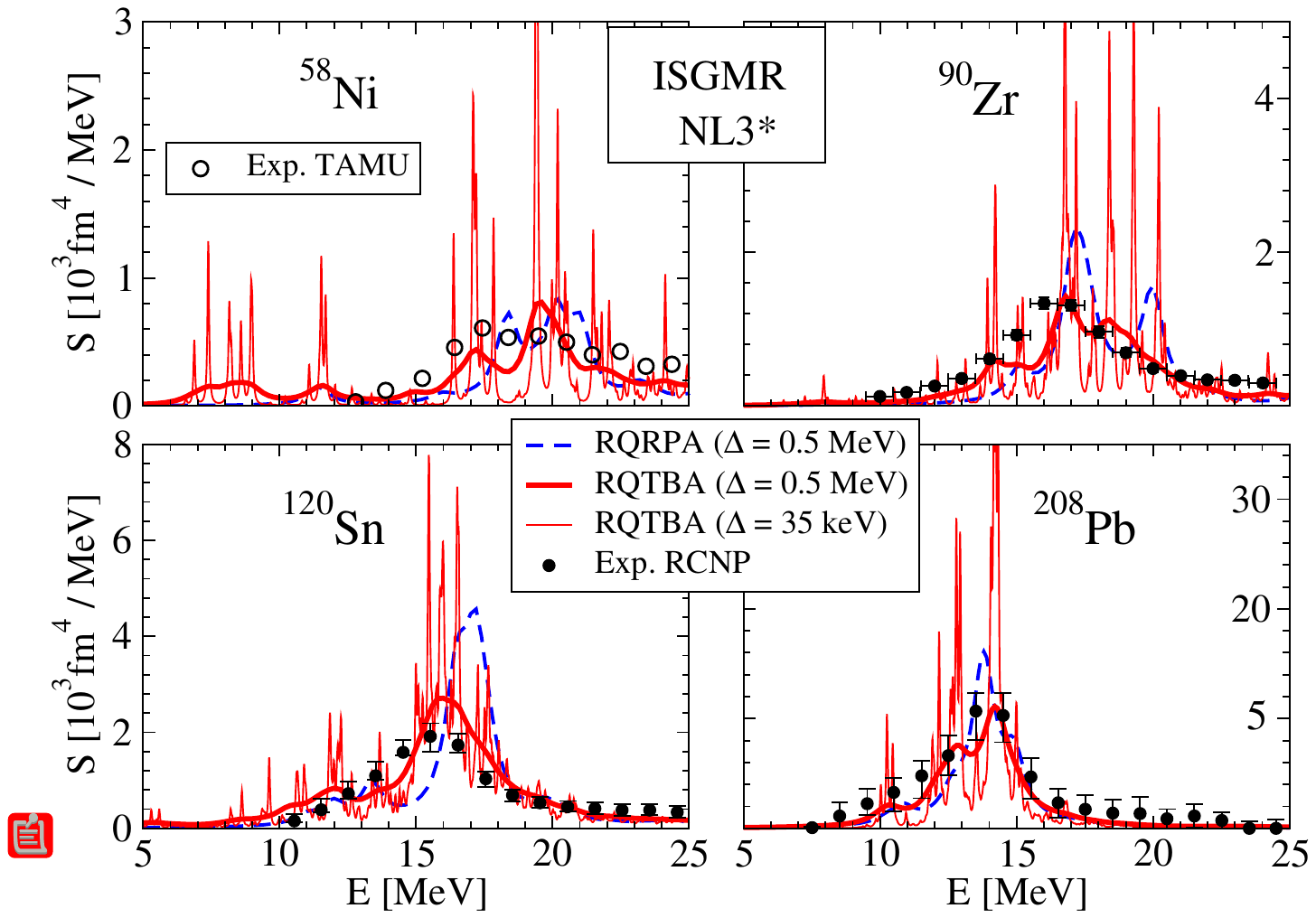}
\end{center}
\vspace{-1cm}
\caption{Isoscalar monopole response of selected medium-mass and heavy nuclei: RQRPA and RQTBA strength distributions compared to experimental data  \cite{Lui2000} ($^{58}$Ni), \cite{Gupta2016} ($^{90}$Zr), \cite{Li2007} ($^{120}$Sn), and \cite{Garg2018} ($^{208}$Pb). RQTBA strength distributions with $\Delta =$ 35 keV and $\Delta =$ 0.5 MeV are shown.}
\label{gmr}%
\end{figure}
\begin{figure*}
\begin{center}
\vspace{-0.5cm}
\includegraphics[width=0.75\textwidth]{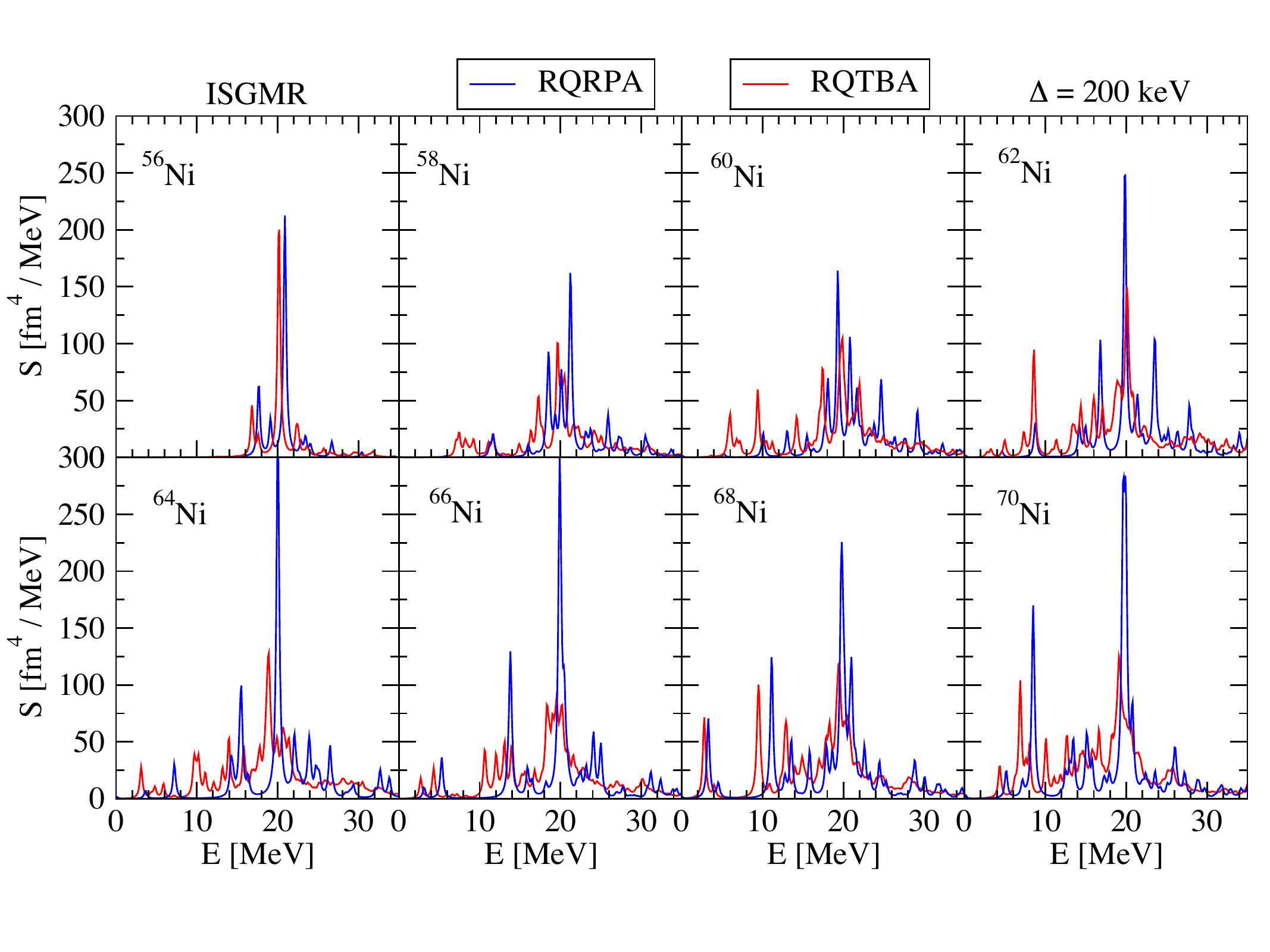}
\end{center}
\vspace{-1cm}
\caption{Isoscalar monopole response of $^{56-70}$Ni isotopes.}
\label{gmr_ni}%
\end{figure*}

Extracting the nuclear compressibility from the ISGMR in finite nuclei, based on the effective interactions derived from the DFT, was extensively discussed in the literature throughout the past decades \cite{Piekarewicz2007,Khan2010,Kvasil2016,Sun2021}. 
It turned out, in particular, that the  DFT parametrizations with commonly accepted $K_{\infty}$ values describing well the ISGMR in $^{208}$Pb notably overestimate the ISGMR centroid in tin isotopes \cite{Garg2018}. The recent calculations of Refs. \cite{Litvinova2023,Li2022} showed that this puzzle could only be resolved by going beyond QRPA, i.e., taking into account the dynamical kernel of BSDE.

In particular, in Ref. \cite{Litvinova2023}, the relativistic nuclear field theory was applied to the description of ISGMR in a variety of nuclear systems to address the long-standing problem of the ISGMR description in a single framework throughout the nuclear chart. 
It was shown that the parameter-free qPVC allows for a simultaneous realistic description of the ISGMR in nuclei of lead, tin, zirconium, and nickel mass regions, which is difficult on the mean-field level. The resolution is already possible in the leading qPVC approximation RQTBA/REOM$^2$.
The calculations employed the finite-range effective meson-nucleon interaction NL3* \cite{NL3star} with $K_{\infty}$ = 258 MeV, which, in combination with the qPVC, has consistently demonstrated the ability to reliably describe many other nuclear structure phenomena.

Fig. \ref{gmr} presents highlights of this work \cite{Litvinova2023}, 
showing the ISGMR in the $^{58}$Ni, $^{90}$Zr, $^{120}$Sn, and $^{208}$Pb nuclei calculated within the self-consistent RQRPA based on Ref. \cite{PaarRingNiksicEtAl2003} in comparison to RQTBA taking into account the qPVC in a parameter-free way \cite{LitvinovaRingTselyaev2008,Litvinova2022}.  The phonon model space included the RQRPA modes of natural parity up to spin $L = 6$ and energy 20-30 MeV, excluding the modes whose B(EL) values are less than 5\% of the maximal one for each spin, i.e., using the same criteria as in the series of earlier calculations, for instance, in Ref. \cite{EgorovaLitvinova2016}. We note here that the subtraction procedure introduced in Ref.  \cite{Tselyaev2013} and becoming the common practice when working with effective interactions, was applied. The approach employed 
monopole pairing forces utilizing the equivalence between the Bardeen-Cooper-Schrieffer (BCS) and Bogoliubov's approaches to the superfluid pairing. The strength of these forces was adjusted in a way to reproduce the energies of the lowest quadrupole states.  

The calculations were performed with $\Delta =$ 35 keV and $\Delta =$ 0.5 MeV. The latter is an adequate choice for comparison to the data in Fig. \ref{gmr}, and the former allows resolving the fine structure of ISGMR, which was the subject of recent studies \cite{Bahini2024}.
It is evident that RQRPA provides a good description of the ISGMR centroid in $^{208}$Pb, while it is positioned noticeably too high in $^{120}$Sn and also a bit too high in $^{58}$Ni and $^{90}$Zr. In RQTBA, we obtained a considerable fragmentation of the RQRPA modes due to the pole structure of the dynamical kernel, including qPVC. In addition, the ISGMR centroids in $^{58}$Ni, $^{90}$Zr, and $^{120}$Sn move to a lower energy to varying degrees, while the one in $^{208}$Pb remains almost intact. In all cases, a definite improvement of the overall ISGMR description is achieved in RQTBA.
Note here that in these calculations, qPVC was included in the leading approximation. While there is still the potential to obtain a better description, here I focus on the leading-order qPVC effect on the ISGMR's centroid. In Ref. \cite{Litvinova2023}, it was conjectured that the major factor influencing its shift is the qPVC to the lowest quadrupole state, which may be more or less collective in nuclei with different locations of their Fermi surfaces relative to the shell closures. The collectivity of the 2$^{+}_1$ states is known to be inversely correlated with their energies. The most pronounced difference in the 2$^{+}_1$ state properties can be observed between $^{208}$Pb, a canonical doubly-magic or closed-shell nucleus, and $^{120}$Sn, a typical neutron mid-shell one.
For instance, E(2$^{+}_1$;$^{208}$Pb) = 4.09 MeV $\gg$ E(2$^{+}_1$;$^{120}$Sn) = 1.17 MeV.  However, these two nuclei are different in many other aspects.
Therefore, instead of comparing them directly, it is more instructive to look at an isotopic or isotonic chain with a similar particle content but significantly varying E(2$^{+}_1$), and a chain of nickel isotopes was chosen for this study.

Fig. \ref{gmr_ni} displays the RQRPA vs RQTBA calculations of ISGMR in even-even $^{56-70}$Ni within the same computational scheme. It provides insights into the ISGMR evolution with the neutron excess, which is already trackable for RQRPA, and simultaneously, the evolution of the leading-order qPVC fragmentation mechanism. Since there are no detailed experimental data on ISGMR in this isotopic chain, the $\Delta = 200$ keV was chosen to capture both the general features of the strength distribution and the fine structure. It is also known that the continuum width of the high-energy peaks in such mid-mass nuclei is about this amount. The main observation from Fig. \ref{gmr_ni} is the relatively constant position of the main peak in RQRPA and the progressive strength increase on the low-energy ISGMR shoulder. This indicates the neutron-skin-related mechanism of the pygmy analog formation in the monopole response, similar to the one reported in Ref. \cite{Gambacurta2019}.  
Further, one can see that the qPVC fragmentation reinforces the enhancement of the low-energy strength and induces a drift of the major peak downward, especially pronounced in $^{60,62,64}$Ni. 
\begin{figure}
\begin{center}
\includegraphics[width=0.48\textwidth]{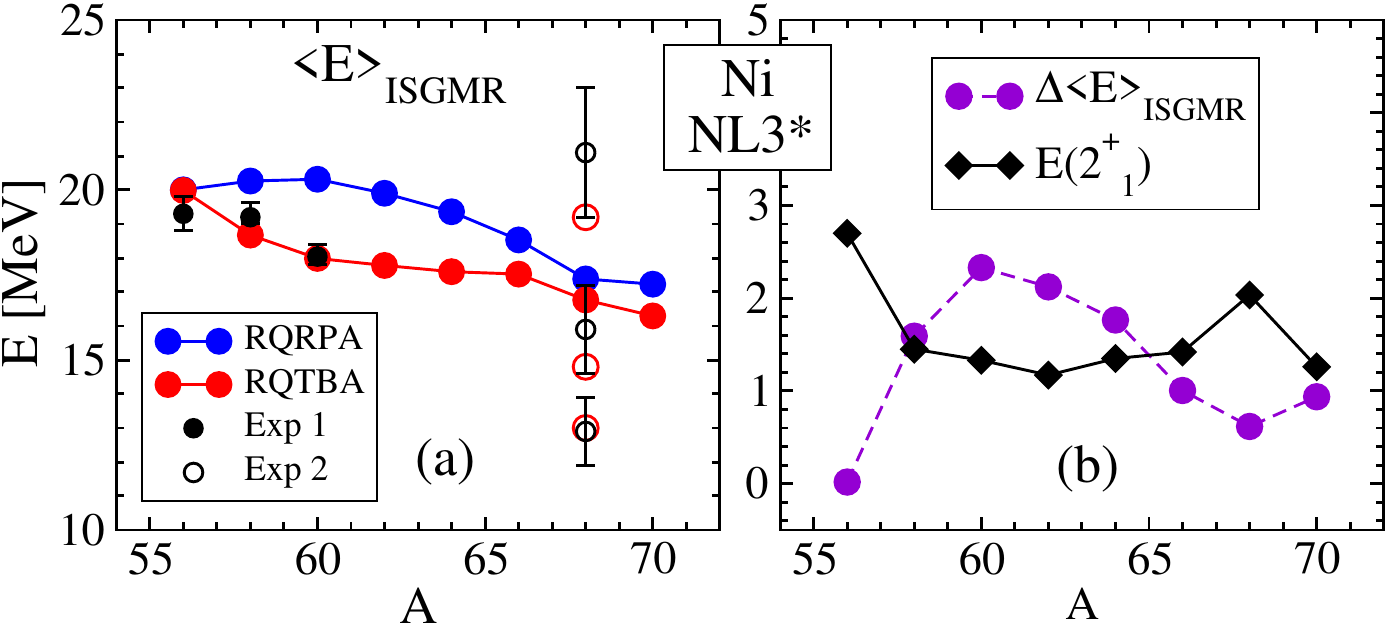}
\end{center}
\vspace{-0.3cm}
\caption{(a) The ISGMR centroids in nickel isotopes $^{56-70}$Ni, compared to data of Refs. \cite{Monrozeau2008} ($^{56}$Ni), \cite{Lui2006} ($^{58,60}$Ni), and \cite{Vandebrouck2014} ($^{68}$Ni).  The three separate peaks above 10 MeV reported in \cite{Vandebrouck2014} are denoted by empty circles with error bars. (b) The downward shifts $\Delta\langle E \rangle_{\text{ISGMR}}$ of the ISGMR centroids obtained in the RQTBA relative to the RQRPA centroids (circles) and the energies of the lowest quadrupole states  E(2$^{+}_1$) (diamonds). The figure is reproduced from Ref. \cite{Litvinova2023}.}
\label{gmr_ni_e0}%
\end{figure}
\begin{figure}
\begin{center}
\vspace{-0.5cm}
\includegraphics[width=0.48\textwidth]{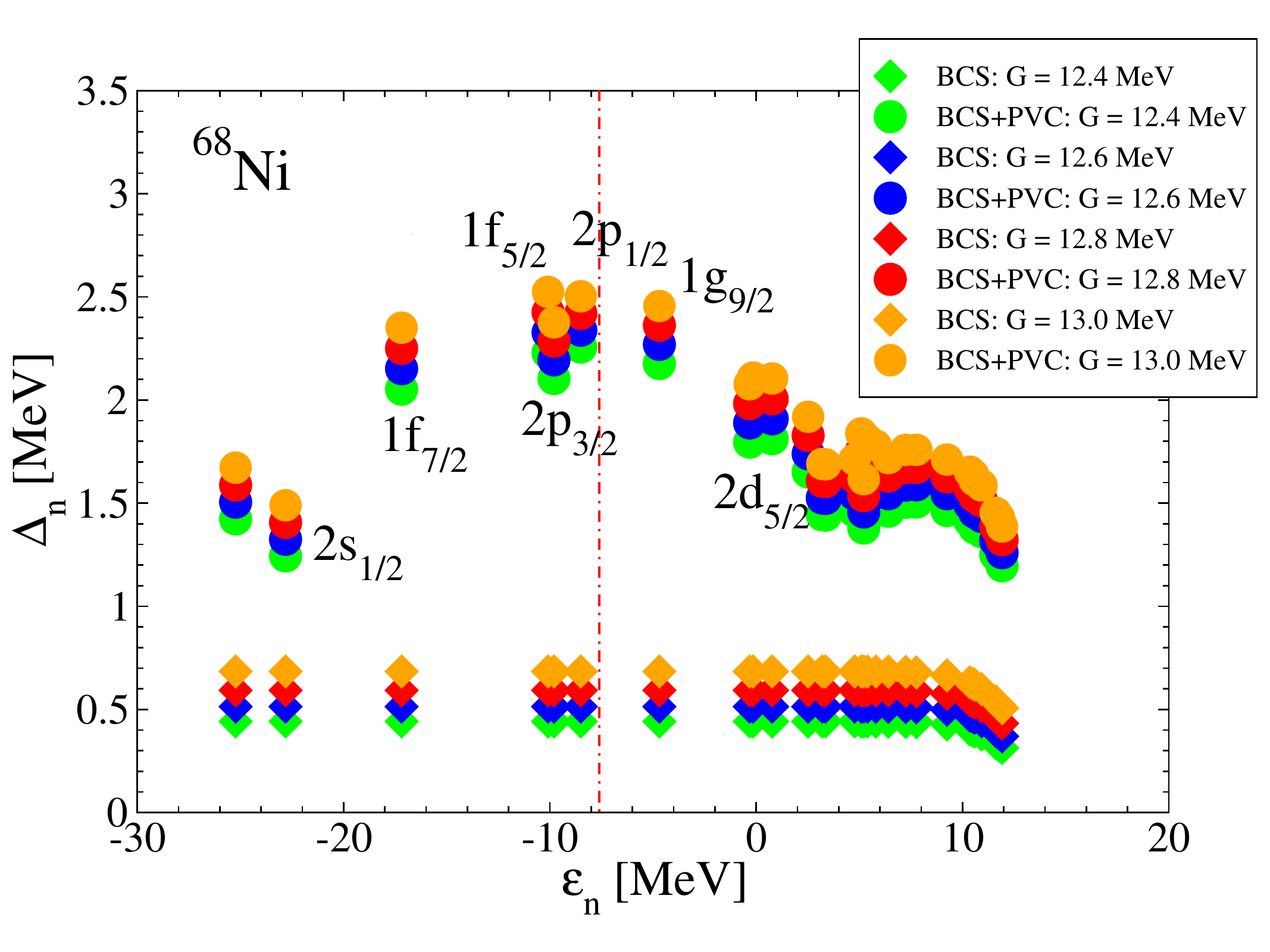}
\end{center}
\vspace{-0.3cm}
\caption{ State-dependent pairing gaps $\Delta_n$ for the nucleonic neutron orbits $|n\rangle$ around the Fermi energy (indicated by the vertical line) in the pure BCS (Hartree-(Fock)-Bogoliubov, HFB) vs BCS (HFB) + PVC approach, for the variable pairing strength parameter $G$. The figure is adapted from Ref. \cite{Litvinova2021}.}
\label{gap}%
\end{figure}

Fig. \ref{gmr_ni_e0} allows one to analyze the conjecture about a correlation between the ISGMR centroid shift due to qPVC and the E(2$^{+}_1$) values.
The centroid was defined as the ratio of the first and zeroth moments $\langle E \rangle_{\text{ISGMR}} = m_1/m_0$ of the obtained ISGMR strength functions $S(E)$ in the energy interval $0\leq E \leq 30$ MeV. Note here that the spurious monopole mode, which can potentially appear at lowest energies and distort these moments, is eliminated due to the self-consistency of the calculations in a sufficiently complete two-quasiparticle basis.
The RQRPA centroid energy indicated by blue symbols in panel (a) of Fig. \ref{gmr_ni_e0} shows a small increase when one moves away from the doubly closed-shell $^{56}$Ni to the open-shell $^{58}$Ni, due to the appearance of superfluidity. After that, a smooth descent occurs with the increase of the neutron number, and the strong subshell closure $N = 40$ causes a kink in $^{68}$Ni. This trend can also be related to the evolution of superfluidity, maximizing the pairing gaps in mid-shell isotopes, and then again suppressed in $^{68}$Ni. 
The red symbols in Fig. \ref{gmr_ni_e0}(a) represent RQTBA results illustrating how the ISGMR centroid energies are changed by qPVC. Their values decrease and show the most pronounced shifts in mid-shell nuclei with a small kink at $^{68}$Ni. For $^{56,60,62}$Ni, the centroids are available from experiment, while for $^{68}$Ni, there are data of Ref. \cite{Vandebrouck2014} for the positions of three peaks above 10 MeV, so that the energies of the corresponding peaks in RQTBA are plotted for comparison. The agreement between the data and RQTBA is reasonable. 

Fig. \ref{gmr_ni_e0}(b) displays the values of the downward shifts $\Delta\langle E \rangle_{\text{ISGMR}}$ of the ISGMR centroids in RQTBA relative to those in RQRPA together with the experimental E(2$^{+}_1$) values used for fitting the pairing gaps. The energies E(2$^{+}_1$) exhibit a well defined countertrend with $\Delta\langle E \rangle_{\text{ISGMR}}$. The increase of the ISGMR centroid shifts corresponds to the decrease of the  E(2$^{+}_1$) values and vice versa. Here we recall that the decrease in  E(2$^{+}_1$) is associated with enhanced collectivity of the quadrupole transitions \cite{Sorlin2002}, which can thus also be linked now to the ISGMR centroid shifts.  
Ref. \cite{Li2022} addressed the ISGMR centroid shifts by qPVC in calculations based on the Skyrme DFT and highlighted the relevance of pairing correlations. The latter are, in turn, directly related to the characteristics of the 2$^{+}_1$ states in medium-heavy nuclei. Indeed, as pointed out above, the calculations of ISGMR presented here were performed with the pairing interaction strength fitted to the E(2$^{+}_1$) values in each nucleus separately. The pairing gaps from such a fitting are by a few keV smaller than those obtained from fitting the odd-even mass staggering \cite{Afanasjev2015}. Furthermore, there is an established qPVC effect on the pairing gaps, suggesting that a noticeable fraction $\sim$50\% of their value in the vicinity of the Fermi surface originates from qPVC \cite{BarrancoBrogliaGoriEtAl1999,Litvinova2021}, and an example from the latter work is given in Fig. \ref{gap}. 
This figure visualizes the role of the qPVC in the pairing gap formation and the sensitivity of the pairing gap values to the parameter $G$, which defines the strength of the static part of the interaction kernel in the particle-particle channel. As in the examples above, the latter kernel was taken in the form of the monopole-monopole interaction given in detail in Ref. \cite{LitvinovaRingTselyaev2008}, while the present study is focused on the qPVC contribution from the dynamical kernel. The parameter $G$ is the only free parameter used for obtaining the pairing and, ideally, will be eliminated in the ab-initio calculations. Here, this parameter was adjusted to reproduce the average experimental value of the pairing gap obtained in the BCS+qPVC calculation, as described in Ref. \cite{Litvinova2021}. One can see from Fig. \ref{gap} that the dynamical qPVC is responsible for more than 50 \% of the pairing gap value. Its contribution is slightly above 50 \% in the peripheral energy regions with respect to the Fermi energy (FE) and increases to 60-70\% for the states close to the FE. Around FE,  the pairing gap exhibits a smooth maximum, which is attributed to the pole structure of the dynamical kernel.

These examples pose two deeply interrelated questions for the implementations of the leading-order qPVC approaches, namely (i) how the pairing gap should be adjusted if it is not part of the DFT fit, and (ii) whether the (R)QRPA phonons are adequate or should they be computed beyond this approximation?
To emphasize the importance of this issue, I consider an example of the sensitivity of the dipole response calculated in RQTBA to the pairing gap values in the next subsection.

\subsection{Electromagnetic dipole response}

The electromagnetic dipole response is the most studied nuclear response probed experimentally with real or virtual photons \cite{GR2001,SavranAumannZilges2013,Bortignon2020,Ishkhanov2021}. In theory, we apply the operator  
\be
F_{1M} = \frac{eN}{A}\sum\limits_{i=1}^Z r_iY_{1M}({\hat{\bf r}}_i) - \frac{eZ}{A}\sum\limits_{i=1}^N r_iY_{1M}({\hat{\bf r}}_i),
\label{FE1}
\ee
where $e$ is the positron charge, $Y_{LM}({\hat{\bf r}})$ are the spherical harmonics, and $N$ and $Z$ are the neutron and proton numbers of the system, respectively, so that  $N + Z = A$. The operator (\ref{FE1}) takes into account the center of mass motion, engaging both the proton and neutron subsystems.
It is regarded as the isovector, or electromagnetic, dipole operator, as opposed to the isoscalar dipole operator associated with a compression mode \cite{GR2001}.

\begin{figure*}
\begin{center}
\includegraphics[width=0.75\textwidth]{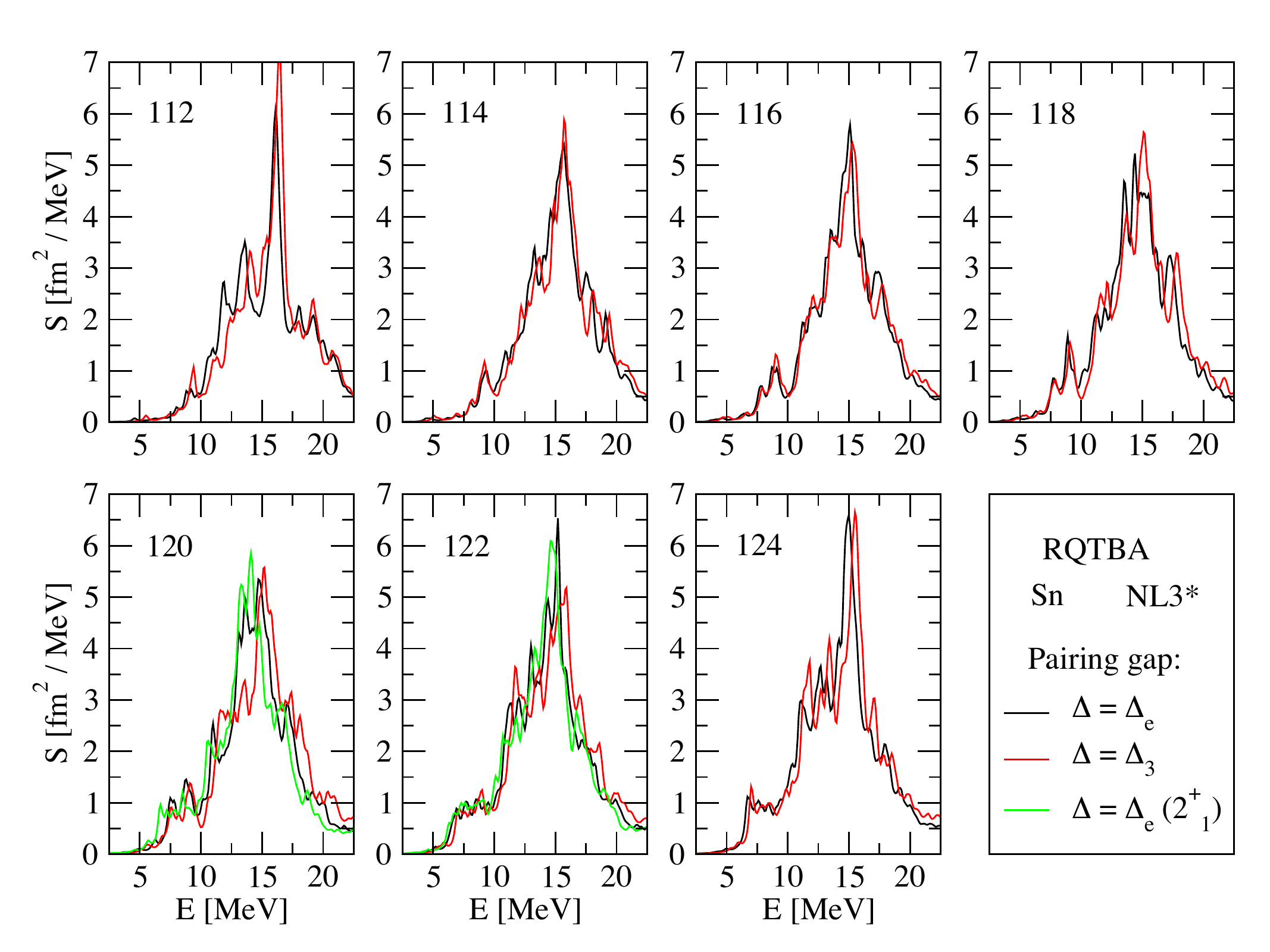}
\end{center}
\vspace{-0.3cm}
\caption{RQTBA dipole response of $^{112-124}$Sn with three different pairing strength values.}
\label{GDR_Sn}%
\end{figure*}

%

The same computational scheme was employed as for the ISGMR described above, for the dipole response of stable even-even tin isotopes. In contrast to earlier studies \cite{LitvinovaRingTselyaev2008}, we looked at a variable pairing gap value fitted to different observables. The rationale is that the value of the pairing gap is not well constrained and is not even well defined in nuclear structure theory. It can be determined uniquely only in a completely ab-initio approach, but also in this case, the pairing gap will be approximate and thus uncertain; only the uncertainties will have a different nature. One of the common practices is to fit the pairing gaps to the odd-even nuclear mass staggering. However, this kind of fit gives only one value for the neutron subsystem and one value for the proton one; moreover, QRPA with such gaps often overestimates the E(2$^{+}_1$) values in spherical nuclei. In the microscopic theory, the pairing gap is, in principle, state-dependent and can be found by solving a nonlinear equation within the BCS approach or its extension, one of which is discussed above.

In this study, we look at the sensitivity of the dipole response to the pairing gap value, which was taken in the BCS approximation but adjusted in different ways, using the freedom offered by the fact that the strength of the pairing interaction is not well constrained. 
The following three methods were applied:  (i) the empirical pairing gap $\Delta_e = 12/\sqrt{A}$ [MeV] (RQRPA ($\Delta_e$), RQTBA ($\Delta_e$)), (ii) the three-point formula applied to the odd-even mass staggering \cite{Afanasjev2015} $\Delta_3$ (RQRPA ($\Delta_3$), RQTBA ($\Delta_3$)), and (iii) the pairing gap reproducing the experimental position of the lowest quadrupole state within RQRPA (RQRPA ($\Delta_e,2^+_1$), RQTBA ($\Delta_e,2^+_1$)). It turns out that for the even-even tin isotopes $^{112-118,124}$Sn, 
$\Delta_3 \approx \Delta_e,2^+_1$, so that only two options were considered for these nuclei.

The RQTBA results for the dipole response of $^{112-124}$Sn spanning a wide energy interval $0 \leq E \leq 25$ MeV with both the pygmy dipole strength and giant dipole resonance (GDR) are displayed in Fig. \ref{GDR_Sn}. 
It illustrates, in particular, the sensitivity of the dipole strength functions to the values of $\Delta_i$, distinguishing the results obtained with different pairing gaps. With the hierarchy $\Delta_3 \geq \Delta_e \geq \Delta_e,2^+_1$, one can observe a similar hierarchy of the GDR main peak location and minor rearrangements in its gross structure. However, an enhanced sensitivity of the low-energy dipole strength, including the pygmy dipole resonance (PDR), can be noticed. While the PDR identification by the specific behavior of transition densities showing the neutron skin oscillation was discussed in detail in Ref. \cite{Markova2025}, here I focus on the issues associated with the uncertainty of the pairing gap. 

\begin{figure}
\begin{center}
\includegraphics[width=0.48\textwidth]{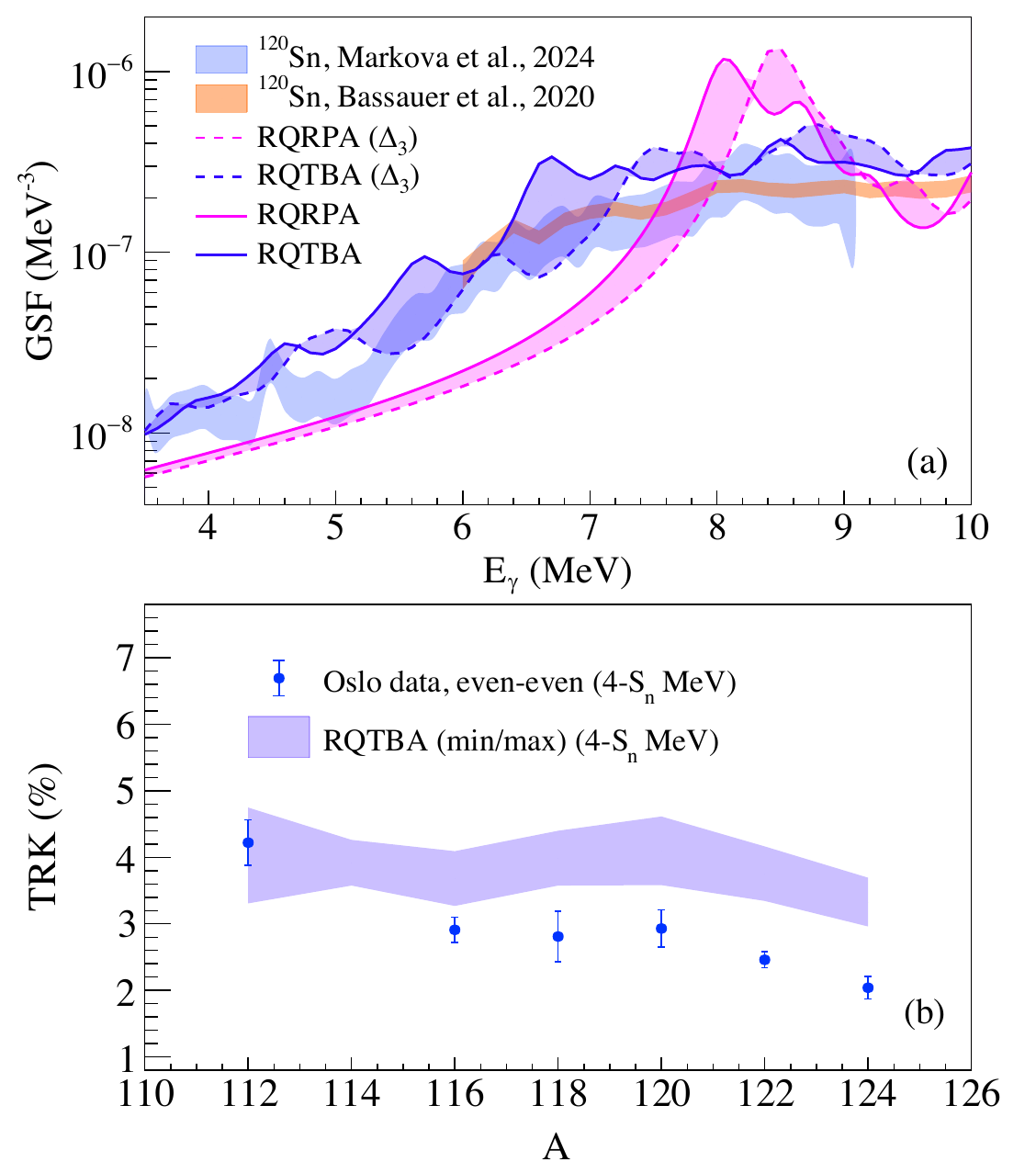}
\end{center}
\vspace{-0.3cm}
\caption{The gamma strength function (GSF) for $^{120}$Sn, proportional to $S_F$ as defined in Ref. \cite{Markova2024,Markova2025} for two different values of the pairing gap: RQRPA and RQTBA, compared to data of Refs. \cite{Markova2024,Bassauer2020} (a);  The total energy-weighted strength, or the fraction of the TRK sum rule below the neutron separation energy $S_n$, in RQTBA for $^{112-124}$Sn, compared to Oslo data \cite{Markova2024}. The band spans the area between  the maximal and minimal sum rule fractions obtained with the range of pairing gaps described in the text (b). The figures are adapted from Ref. \cite{Markova2025}. }
\label{PDR_Sn}%
\end{figure}

In this approach, nuclear pairing is expressed in the non-monopole response only indirectly, namely, via the energies of Bogoliubov quasiparticles $E_i = \sqrt{\varepsilon_i^2 + \Delta_i^2}$, where $\varepsilon_i$ are the mean-field nucleonic energies on the orbits $|i\rangle$, counted from the chemical potential, and $\Delta_i$ are the state-dependent pairing gaps \cite{RingSchuck1980}, because the superfluidity is supposed to occure exclusively via the monopole Cooper pairs. Another indirect effect is associated with the change of the phonon characteristics, especially the low-lying ones. The influence of the E(2$^{+}_1$), which is mostly sensitive to the change in pairing, on the monopole response was discussed in the previous subsection, and here there is a similar effect. The sensitivity to changes in other phonon modes is weaker but also present.
Fig. \ref{PDR_Sn}(a) presents an enhanced view of the low-energy part of the dipole spectrum for $^{120}$Sn computed in RQTBA with the $\Delta_e,2^+_1$ and  $\Delta_e$ pairing gap values obtained for this nucleus. This representation allows a better assessment of the sensitivity of the low-energy dipole strength distribution to the pairing gap value. As the peak locations are determined by the poles at $E_{ij} = E_i+E_j$ (RQRPA) or $E_{ij} + \omega_{\nu}$ (RQTBA) shifted by the effective interaction, it is understandable that the low-energy peaks are more affected by pairing than the high-energy ones. With the average value of $\Delta_i \approx 1$ MeV, the shift of the major peaks below 10 MeV is of the order of 1 MeV. Fig. \ref{PDR_Sn}(b) provides an assessment of the fraction of the Thomas-Reiche-Kuhn (TRK) sum rule below the neutron separation energy, across the stable tin isotopes, compared to data. Both the RQRPA and RQTBA energy-weighted strengths are given in the form of bands spanning the area between the smallest and largest values obtained with the minimal $\Delta_e,2^+_1$ and maximal $\Delta_3$ pairing gaps.
This demonstrates that the uncertainties in the pairing correlations propagate significantly enough in the low-energy dipole strength to cause critical uncertainties of the neutron capture rates, which are to be extracted from these spectra for nucleosynthesis modeling.

Thus, these results call for a refined procedure for determining the effective interaction in the pairing channel and accounting for the qPVC in the gap equation and in obtaining the phonon characteristics in the approach of progressive inclusion of complex configurations $npnh$ with growing degree of complexity $n$. The internal paradox of this approach is that it is designed to enable a refinement of the calculated spectra, but it introduces uncertainties stemming from the choice of a calculation scheme, identifying and ordering quantitatively important nucleonic correlations, and coherently adjusting the interaction parameters in the $ph$ and $pp$ channels.  To resolve these difficulties, it is particularly desirable to accurately bridge the definitions and relationships between the bare and effective Hamiltonians.

\section{Summary and outlook}

An overview of the modeling of the nuclear isoscalar monopole and isovector dipole responses in a beyond-QRPA approach in the leading order of qPVC, taking into account the major effects on emergent collectivity on the spectral strength formation, is presented. In the monopole sector, the focus is placed on the unified and predictive description of both the ISGMR's width and its centroid across the nuclear chart. The centroid problem was analyzed in detail because of its unique relation to the nuclear compressibility modulus.
Systematic calculations of the isoscalar monopole response for nickel isotopes revealed the central role of the coupling between the ISGMR and the low-energy quadrupole states in the placement of the ISGMR centroids. It was shown that, although it generally depends on the nuclear compression modulus associated with a particular parameter set, it is further fine-tuned naturally by the coupling of the ISGMR to the low-energy phonons, mostly those of quadrupole character. This coupling causes spreading and an overall shift of the ISGMR centroid down with respect to its value obtained in the RQRPA, which can amount to 1-2 MeV and which is more pronounced in softer mid-shell nuclear species, in agreement with data. This allows one to reconcile the nuclear and neutron star data, suggesting that the same DFT-based interaction can consistently describe the ISGMR beyond QRPA in soft nuclei and be applied to nuclear matter. This result is rather robust, being obtained with $K_{\infty}$ = 258 MeV associated with the NL3$^{\ast}$ interaction used in this work and with $K_{\infty}$ = 226 MeV of one of the Skyrme interactions \cite{Li2022}. Although the covariant DFT produces a stiffer EOS, the reasonable description of the ISGMR 
obtained with $K_{\infty}$ = 258 MeV 
further supports the use of CDFT-based approaches in nuclear matter calculations and suggests that the latter should also explore beyond-mean-field many-body solutions.

Furthermore, the dipole response is also found to be sensitive to the reproducibility of characteristics of the lowest quadrupole phonons. This was traced back to the pairing gap values and, more generally, to the strength of the nuclear pairing interaction. An RQTBA study of the stable even-even tin isotopes provides a quantitative assessment of the sensitivity of the dipole response, enhanced in the low-energy region, to the superfluid pairing gap values. On one hand, the self-consistent implementations of qPVC can benefit from this, as the propagation of the pairing gap uncertainties is well controlled and can thus provide uncertainties on the nuclear strength functions. On the other hand, this study opens an avenue to reduce the uncertainties on the pairing gaps by refining the ground state description beyond BCS (HFB) and the phonon characteristics beyond QRPA.

\section*{Acknowledgement}
I thank Maria Markova for adapting Fig. \ref{PDR_Sn} originally appearing in Ref. \cite{Markova2025}, co-authored by us and P. von Neumann-Cosel.
This work was supported by the GANIL Visitor Program and US-NSF Grants PHY-2209376 and PHY-2515056.
%
\bibliography{Bibliography_Jun2024}
\end{document}